\documentclass[aip,amsmath,amssymb,reprint]{revtex4-1}
\newcommand{\ket}[1]{\left\vert #1 \right\rangle}
\newcommand{\bra}[1]{\left\langle #1 \right\vert}

\usepackage{graphicx}
\usepackage{epsfig}
\usepackage{amsmath}
\usepackage{amssymb}

\usepackage{dcolumn}
\usepackage{bm}

\usepackage[utf8]{inputenc}
\usepackage[T1]{fontenc}
\usepackage{mathptmx}
\usepackage{etoolbox}
\makeatletter
\def\@email#1#2{%
 \endgroup
 \patchcmd{\titleblock@produce}
  {\frontmatter@RRAPformat}
  {\frontmatter@RRAPformat{\produce@RRAP{*#1\href{mailto:#2}{#2}}}\frontmatter@RRAPformat}
  {}{}
}%
\makeatother 
\begin{document}
%----------------------------------------------------
%
\bibliographystyle{apsrev}
\preprint{version: \today}
%
%\title{Zombie cats on the quantum-classical frontier}
\title{Zombie cats on the quantum-classical frontier: Wigner-Moyal and semiclassical limit dynamics of quantum coherence in molecules}
\author{Austin T.~Green and Craig C.~Martens}
\affiliation{University of California, Irvine, California 92697-2025}
\email[email: ]{cmartens@uci.edu}
\date{\today}
%
%----------------------------------------------------
\begin{abstract}

In this paper, we investigate the time evolution of quantum coherence---the off-diagonal elements of the density matrix of a multistate quantum system---from the perspective of the Wigner-Moyal formalism.  This approach provides an exact phase space representation of quantum mechanics.  We consider the coherent evolution of nuclear wavepackets in a molecule with two electronic states. For harmonic potentials, the problem is exactly soluble for both fully quantum and semiclassical descriptions.   We highlight serious  deficiencies of the semiclassical treatment of coherence for general systems and illustrate how even qualitative accuracy requires higher terms in the Moyal expansion to be included.  The model provides an experimentally relevant example of a molecular Schr\"{o}dinger's cat state.  The alive and dead cats of the exact two state quantum evolution collapses into a ``zombie'' cat in the semiclassical limit---an averaged behavior, neither alive nor dead, leading to significant errors. The inclusion of the Moyal correction restores a faithful simultaneously alive and dead representation of the cat that is experimentally observable.

\end{abstract}
%
%----------------------------------------------------
\maketitle
%----------------------------------------------------
%

\section{Introduction}

The classical limit of quantum mechanics has been a topic of intense interest since the earliest days of quantum theory \cite{dira57,feyn65,jamm66}. Despite this long history, quantum-classical correspondence is still an active area of research with both conceptual challenges and practical applications.  Aside from its fundamental interest, this problem has direct relevance to quantum information theory, coherent energy transfer, quantum biology, solar energy, and other areas of current importance.  

Semiclassical mechanics falls in the borderland between the classical and quantum worlds.   It provides both a philosophical vantage point for the interpretation of quantum theory and the foundation for an array of computational methods for calculating quantum properties using classical quantities \cite{chi91,hell18}.  

Molecules typically reside along the border between the quantum and classical worlds, with one foot on each side.  The light mass of electrons results in their molecular properties being highly quantum mechanical, while the much heavier nuclei can often be described in classical terms.  The Born-Oppenheimer approximation \cite{born54} provides a theoretical framework for taking advantage of the different masses and resulting timescales for electronic and nuclear motion by making an adiabatic separation of electronic and nuclear dynamics.  This allows a mixed quantum-classical framework, where classical molecular dynamics describes nuclei following classical trajectories that evolve under forces derived from the configuration dependence of the adiabatic ground state quantum energy.  

Things get more interesting---and difficult---when the electronic-nuclear timescale separation breaks down and electronic excitation and transition accompany the nuclear motion.  A rigorous fully quantum description involves the coherent and entangled evolution of coupled electrons and nuclei.  Here, a coupled multicomponent set of multidimensional nuclear quantum wavepackets are propagated using a convenient electronic basis.  Although efficient methods have been developed that allow multidimensional systems to be treated efficiently \cite{beck00}, for large molecules this approach eventually becomes prohibitively expensive and additional approximations must be made.  

The development of semiclassical and mixed quantum-classical methods for simulating molecular dynamics with nonadiabatic transitions an active area of research.  Much effort is focused on developing theories and methods that describe the quantum dynamics of molecular systems using an underlying classical trajectory framework.  Such ongoing programs include trajectory surface hopping \cite{tull90,Subotnik:2016ch}, full multiple spawning \cite{tjm96a,tjm96b,benn98,hack01}, semiclassical initial value representation \cite{sun97,xion98,sun99,mill01,moix08}, quantum hydrodynamics \cite{lopr99,bitt00a,bitt00b,lopr02,bous11}, mapping Hamiltonian approaches \cite{Meyer:1979vu,bone01,stockthossacp05,Nassimi:2010ht,Nassimi:2009dn}, linearized propagation \cite{Bonella:2005hj,Bonella:2005wk,Dunkel:2008dv,Huo:2010p15441,Huo:2011fq}, Bohmian dynamics \cite{Curchod:2013cq}, symmetrical windowing of quasiclassical trajectories \cite{Cotton:2013fa}, ring polymer methods \cite{Richardson:2013jm,Ananth:2013fp}, exact factorization \cite{vine22}, trajectory-guided Gaussian wavepackets \cite{White:2016ic,Humeniuk:2016fj}, and quantum-classical Wigner function-based approaches \cite{mart97,dono98,dono00a,dono00b,dono02b,kapr99,Bonella:2010hz,Kapral:2006p6455}.  In general, these methods hold great promise because they allow the essential quantum features of complex systems to be incorporated within the computationally less demanding and intuitively appealing framework of classical molecular dynamics.  

The coupled electronic-nuclear dynamics of a molecule with two electronic states can be described by the evolution of the $2 \times 2$ density matrix $\hat{\rho}_{ij}$ $i,j = 1,2$.  Here, the elements are nuclear operators and the indices correspond to the chosen electronic basis.  The diagonal elements $\hat{\rho}_{jj}$ correspond to the populations of the electronic states, while the off-diagonal element $\hat{\rho}_{ij}=\hat{\rho}_{ji}^\dagger$ corresponds to the coherence between the electronic basis states. 

Electronic state populations have clear classical analogues: phase space probability densities, which can be modeled by ensembles of classical trajectories.  Quantum coherence between electronic states, on the other hand, is manifestly quantum mechanical with no direct classical analogue.  It is fair to say that most existing methods for modeling classical-limit molecular dynamics with electronic transitions have conceptual inconsistencies in dealing with coherence that translate into challenges and errors in their practical implementation.  

In this paper, we investigate exact quantum and approximate semiclassical coherence in molecules with multiple electronic states from a quantum phase space perspective.  The Wigner-Moyal representation of quantum mechanics provides a particularly useful perspective on the connections between quantum and classical mechanics as well as a convenient theoretical framework for developing quantum-classical methods \cite{wign32,moya49,groe46,zach05}.  The formalism provides an exact phase space representation of the states and operators of quantum systems. Quantum effects emerge \emph{via} a noncommutative generalization of the commutative algebra of classical functions \cite{zach05}.  In this representation, the nuclear operator $\hat{A}$ representing an observable becomes the corresponding phase space function $A(q,p)$.  Density matrix elements $\hat{\rho}_{ij}$ become phase space functions $\rho_{ij}(q,p,t)$ as well.  

 A great deal of mathematical literature focuses on the problem of deformation quantization---the extension of Hamiltonian dynamics into the quantum realm as a function of the small parameter $\hbar$.  In physics and chemistry, this field has focused on phase space representations of quantum states, where higher order terms in the Moyal expansion are interpreted as a quantum corrections to the underlying scaffolding of classical trajectories carrying semiclassical phase interference effects.  Here, we address a more subtle and less appreciated failure of semiclassical theory to describe the coherence between states for even simple systems.

Using the Wigner Moyal formalism, we highlight significant errors in the description of coherence in the semiclassical limit, and illustrate how accuracy is restored when higher order terms in the Moyal expansion (described below) are included.  The general principles are illustrated for the specific case of an initially prepared quantum vibrational coherence between two electronic states. Such excitations form the conceptual framework for time-dependent theoretical spectroscopy \cite{muka95,hell18}, and modern experimental techniques allow time-dependent coherences to be produced in the laboratory as well.  This corresponds to  a so-called Schr{\"o}dinger's cat state, studied previously in quantum physics and optics \cite{schr35,schl01,wine2013}  and realized experimentally in molecular systems (see, e.g.,~\cite{kowa2015,buck2016} for recent examples).  Aside from its fundamental interest, the preparation, evolution, and decay of quantum coherence forms the basis of many applications in molecular dynamics and spectroscopy (see, e.g.,~\cite{muka95,nitz06,hell18}). 

We treat a very simple but nontrivial model of two electronic states with one-dimensional displaced harmonic vibrational potentials describing the nuclear dynamics.  The potentials are shifted with respect to each other and possess different frequencies in general.  It is well-known that quantum dynamics on a \emph{single} harmonic potential in the Wigner representation corresponds to purely classical evolution of the Wigner function, as nonclassical corrections vanish if anharmonicity is absent \cite{tann07,hell18}.  The \emph{coherence} between displaced oscillator states, however, is \emph{not} purely classical, as we will see below.  Despite this, the exact dynamics of the electronic coherence in the Wigner representation can be analyzed analytically for the two-state harmonic system.  

\section{Wigner-Moyal Representation}

We now briefly review the Wigner-Moyal representation of quantum mechanics \cite{zach05}.  This approach associates classical-like functions of the canonical phase space coordinates $(q,p)$ with quantum mechanical operators, and generalizes the classical Poisson bracket that underlies classical Hamiltonian dynamics to the \emph{Moyal bracket} in the quantum case.  

Operators corresponding to quantum observables are represented by phase space functions called Weyl symbols, given by \cite{zach05}
\begin{equation}
	A_W(q,p) = \int \bra{q+\tfrac{y}{2}} \hat{A} \ket{q-\tfrac{y}{2}} e^{-i p y/\hbar} \, dy.
\end{equation}
In Hilbert space formulations of quantum mechanics, the state of a system is represented most generally by the density operator $\hat{\rho}$.  In the Wigner-Moyal representation, this corresponds to a phase space function called the Wigner function $\rho_W(q,p)$ \cite{zach05}:
\begin{equation}
	\rho_W(q,p) = \frac{1}{2 \pi \hbar} \int \bra{q+\tfrac{y}{2}} \hat{\rho} \ket{q-\tfrac{y}{2}} e^{-i p y/\hbar} \, dy.
\end{equation}
For a pure state represented by the wavefunction $\psi(q)$, the Wigner function becomes
\begin{equation}
	\rho_W(q,p) = \frac{1}{2 \pi \hbar} \int \psi(q+\tfrac{y}{2}) \psi^*(q-\tfrac{y}{2}) e^{-i p y/\hbar} \, dy.
\end{equation}

The fundamental ``quantumness'' of quantum mechanics is manifested in the noncommutativity of the operators representing physical observables.  In the Wigner-Moyal representation, this property is captured by the role of the \emph{Moyal} or \emph{star} product, a noncommutative generalization of the limiting commutative product of functions in phase space \cite{zach05}.  In particular, the Weyl symbol of the product of operators $\hat{A}$ and $\hat{B}$ becomes the star product of the individual Weyl symbols:
\begin{equation}
(AB)_W(q,p) = A_W(q,p) \star B_W(q,p).
\end{equation}
We define the bidirectional differential operator $
\overleftrightarrow{\Lambda} = \overleftarrow{\partial}_q \overrightarrow{\partial}_p - \overleftarrow{\partial}_p \overrightarrow{\partial}_q$, where the direction of the over arrows indicates the direction (left or right) that the derivative operators act.  In this notation, the classical Poisson bracket \cite{gold80} can be written as $\{A,B\} = A \overleftrightarrow{\Lambda} B$.  The star product of the Weyl symbols $A_W$ and $B_W$ is then defined as
\begin{equation}
A_W \star B_W = A_W e^{\frac{i \hbar}{2} \overleftrightarrow{\Lambda}} B_W.
\end{equation}
In practice, the star product can be evaluated by expanding $\exp(i \hbar \overleftrightarrow{\Lambda}/2)$ in a power series in $\hbar$, which we refer to as the Moyal series:
\begin{multline}
A_W \star B_W = \\ A_W B_W + \frac{i \hbar}{2} A_W \overleftrightarrow{\Lambda} B_W - \frac{ \hbar^2}{8} A_W \overleftrightarrow{\Lambda}^2 B_W + O(\hbar^3).
\end{multline}
It should be kept in mind, however, that series expansions in the ``small parameter'' $\hbar$ are typically asymptotic series with problematic convergence properties \cite{hell18}.  

The density operator $\hat{\rho}(t)$ describes the time evolution of the state of a quantum mechanical system.  It obeys the quantum Liouville equation, given by
\begin{equation}
	i \hbar \frac{d \hat{\rho}}{dt} = [\hat{H},\hat{\rho}]=\hat{H}\hat{\rho}-\hat{\rho}\hat{H},
\end{equation}
where $\hat{H}$ is the system Hamiltonian.  We can express the Liouville equation in terms of Weyl symbols and the star product.  Performing the Weyl transform of both sides  gives the \emph{exact} equation of motion for $\rho_W(q,p,t)$:
\begin{equation} \label{eq:moyalliouv}
\frac{\partial \rho_W}{\partial t} = \frac{1}{i \hbar} (H_W \star \rho_W - \rho_W \star H_W) \equiv \{\{H_W,\rho_W\}\}.	
\end{equation}
This defines the Moyal bracket of the two Weyl symbols $A_W$ and $B_W$:
\begin{equation} \label{eq:weylbrac}
	\{\{A_W,B_W\}\} = \frac{1}{i \hbar} (A_W \star B_W - B_W \star A_W).
\end{equation}
An expansion of Eq.~(\ref{eq:weylbrac}) in $\hbar$ has as its leading term the classical Poisson bracket of the Weyl symbols $A_W$ and $B_W$: $\{\{A_W,B_W\}\}  =  \{A_W,B_W\} +  O(\hbar^2)$.  
The classical limit of Eq.~(\ref{eq:moyalliouv}) thus gives the familiar classical  Liouville equation,
\begin{equation}
	\frac{\partial \rho_W}{\partial t} = \{H_W,\rho_W\}.
\end{equation}
In this limit, the Wigner function evolves purely classically. 

\section{Two State System}

We now consider the Wigner-Moyal dynamics and its semiclassical limit for the particular case of nuclear dynamics in a system with two electronic states.  Here, the phase space representation is introduced by a \emph{partial} Wigner-Moyal transform over nuclear coordinates, retaining a two state representation for electronic degrees of freedom \cite{mart97,dono98,kapr99}.  

We focus in particular on the \emph{coherence} between the two states, represented by $\hat{\rho}_{12}$, the off-diagonal matrix element of the density operator in the electronic basis.  

The Hamiltonian $\hat{H}$ and density operator $\hat{\rho}$ describing nuclear dynamics on two  electronic states are given by the $2 \times 2$ matrices of nuclear operators:
\begin{equation} \label{eq:hmatrixd}
  \hat{H}=\left(
    \begin{array}{c}
      \hat{H}_{11} \ \ \hat{H}_{12}  \\
      \hat{H}_{21} \ \ \hat{H}_{22}
    \end{array} \right)
   \,\,\,\,\,\,\,\, \mathrm{and} \,\,\,\,\,\,\, \hat{\rho}=\left(
    \begin{array}{c}
      \hat{\rho}_{11} \ \ \hat{\rho}_{12}  \\
      \hat{\rho}_{21} \ \ \hat{\rho}_{22}
    \end{array} \right),
 \end{equation}
respectively.  We treat the case of coherent dynamics in the absence of interstate coupling.  Here, $\hat{H}_{12}=0$, and the equation of motion for the coherence $\hat{\rho}_{12}$ decouples from the diagonal elements $\hat{\rho}_{jj}$:
\begin{equation} \label{eq:quantcohliouv}
i \hbar \frac{d \hat{\rho}_{12}}{dt} = \hat{H}_{11} \hat{\rho}_{12} - \hat{\rho}_{12} \hat{H}_{22}.	
\end{equation}
In the Wigner-Moyal representation, this becomes (dropping hereafter the subscript ``$W$'' for notational simplicity):
\begin{equation} \label{eq:wigmoycoh}
i \hbar \frac{\partial \rho_{12}}{\partial t} = H_{11} \star \rho_{12} -\rho_{12} \star H_{22},
\end{equation}
or, for systems with Hamiltonians of the form $H_{jj} = \frac{p^2}{2 m} + U_j(q)$,
\begin{equation} 
\frac{\partial \rho_{12}}{\partial t} = \{H_o,\rho_{12}\} - i \omega \rho_{12} + \frac{i\, \hbar^2 }{8} \omega''\frac{\partial^2 \rho_{12}}{\partial p^2} + \cdots \, ,
\end{equation}
where $H_o = (H_{11}+H_{22})/2$ is the average Hamiltonian, $\omega(q) = (U_1(q)-U_2(q))/\hbar$ is the difference frequency, and $\omega'' = d^2 \omega(q)/dq^2$.  

We now specialize to the case of two one-dimensional harmonic diagonal Hamiltonians $H_{jj}(q,p) = \frac{p^2}{2 m} + \frac{1}{2} m \,\Omega_j^2 (q-Q_j^e)^2 + E_j^e$ ($j=1,2)$.  Here, $m$ is the mass, $\Omega_j$ are the harmonic frequencies, and $E_j^e$ are the vertical energies of the states at the potential minima $Q_j^e$.  In general, these harmonic potentials have different frequencies and their equilibrium positions are shifted with respect to one another.  For such systems, the Moyal series terminates, and the exact equation for the coherence Wigner function $\rho_{12}(q,p,t)$ is given by the first three terms of the previous expansion:
\begin{equation} \label{eq:pdeexact}
\frac{\partial \rho_{12}}{\partial t} = \{H_o,\rho_{12}\} - i \omega \rho_{12} + \frac{i \hbar^2 \omega''}{8} \frac{\partial^2 \rho_{12}}{\partial p^2} \, ,
\end{equation}
where now $\omega''$ is a constant.  

It is important to note that a nonclassical term appears in the evolution of the coherence even for harmonic systems if $\Omega_1 \neq \Omega_2$.  This is contrary to the commonly held conviction that the quantum dynamics of harmonic systems are always ``classical''.  The Weyl symbol of the commutator of a harmonic Hamiltonian with the density operator indeed reduces to the classical Poisson bracket, but the fact that $\hat{H}_{11} \neq \hat{H}_{22}$ in Eq.~(\ref{eq:quantcohliouv}) leads to additional physics in the evolution of the manifestly nonclassical coherence.  

\subsection{Moyal Dynamics}  Despite the presence of the higher order nonclassical term,  Eq.~(\ref{eq:pdeexact}) is exactly soluble for Gaussian initial conditions.  To proceed, we use a thawed Gaussian approach, similar to that of Heller for studying wavepacket dynamics of the Schr\"odinger equation \cite{hell75}, but here applied in phase space.  In particular, we use the ansatz
\begin{multline} \label{eq:ansatz}
	\rho_{12}(q,p,t) \\= \exp [ -a (q-Q)^2 - b (p-P)^2  + c (q-Q)(p-P) \\ + u (q-Q) + v(p-P) + w ],
\end{multline}
where the parameters $(a,b,c,Q,P,u,v,w)$ are all functions of time.  This expression is substituted into Eq.~(\ref{eq:pdeexact}) and, by equating powers of $(q-Q)^m (p-P)^n$ on the left and right sides, a set of ordinary differential equations  is derived for $(a,b,c,Q,P,u,v,w)$. For a given set of initial conditions, the solution of these coupled differential equations gives the exact time evolution of $\rho_{12}(q,p,t)$.  

Analytic solutions can be found for the resulting equations.  The analysis is lengthy, and we only highlight a few key aspects of the results here.  We focus in particular on the evolution of the center of the Gaussian, $(Q,P)$, which can be interpreted as a trajectory in phase space.  

For the full Wigner-Moyal dynamics, the differential equations for $(Q,P,u,v)$ are coupled to each other but are independent of the parameters $(a,b,c,w)$.  In particular, we find
\begin{equation} \label{eq:qp}
\dot{Q} = \frac{P}{m}	\,\,\,\,\,\,\,\,\,\,\,\,\, \dot{P} = - U_o'(Q) - \frac{i \hbar^2}{4} \omega'' v	
\end{equation}

\begin{equation} \label{eq:uv}
\dot{u} = - i \omega'(Q) + U_o'' v	\,\,\,\,\,\,\,\,\,\,\,\,\, \dot{v} = -\frac{1}{m} u	,
\end{equation}
where $U_o(Q) = \frac{1}{2}\left(U_1(Q)+U_2(Q)\right)$ is the average potential, which figures importantly in our analysis below.  For the harmonic system, $U_o''$ and $\omega''$ are constants and $U_o'(Q)$ and $\omega'(Q)$ are both linear functions of $Q$. Equations (\ref{eq:qp}) and (\ref{eq:uv}) can therefore be solved analytically by linear algebraic methods.  The results can be expressed as
\begin{equation} \label{eq:qpdex}
Q(t) =	\frac{1}{2}(Q_1(t)+Q_2(t))  \,\,\,\,\,\,\,\,\,\,\,\,\, P(t) =	\frac{1}{2}(P_1(t)+P_2(t))
\end{equation}
\begin{equation}\label{eq:uvdex}
u(t) =	\frac{i(P_1(t)-P_2(t))}{\hbar} \,\,\,\,\,\,\,\,\,\,\,\,\, v(t) =-\frac{i(Q_1(t)-Q_2(t))}{\hbar},
\end{equation}
where the quantities $(Q_1(t),P_1(t))$ and $(Q_2(t),P_2(t))$ are \emph{distinct} classical harmonic oscillator trajectories evolving on state 1 and 2, respectively, with the common initial conditions $(Q(0),P(0))$, 

\begin{equation} \label{eq:qj}
Q_j(t) = Q_j^e + (Q(0)-Q_j^e)\cos(\Omega_j t) + \frac{P(0)}{m \Omega_j} \sin(\Omega_j t)
\end{equation}
\begin{equation} \label{eq:pj}
	P_j(t) = P(0)\cos(\Omega_j t) - m \Omega_j (Q(0)-Q_j^e) \sin(\Omega_j t)
\end{equation}
 for $(j=1,2)$.  The Gaussian parameters $(Q(t),P(t))$ are then the arithmetic means of these independent state-specific trajectories.  The parameters $u(t)$ and $v(t)$ are purely imaginary, and are proportional to the differences of the state specific trajectories.  (The fact that $u$ and $v$ are purely imaginary confirms that $(Q,P)$ is indeed the phase space center of the modulus of $\rho_{12}$.)  

The coupling resulting from the $\hbar^2$-dependent Moyal correction term in Eq.~(\ref{eq:qpdex}) leads to the appearance of the two distinct trajectories $(Q_1(t),P_1(t))$ and $(Q_2(t),P_2(t))$, each associated with one of the two electronic states, comprising the average $(Q(t),P(t))$ and difference $(u(t),v(t))$. The exact Wigner-Moyal evolution is illustrated schematically in Fig.~\ref{fig:fig1} a).    

\subsection{Semiclassical Dynamics}  In the semiclassical limit, the Liouville equation in the Wigner representation reduces to the simpler partial differential equation,
\begin{equation} \label{eq:pdesc}
\frac{\partial \rho_{12}}{\partial t} = \{H_o,\rho_{12}\} - i \omega \rho_{12}	.  
\end{equation}
A solution can again be obtained using the phase space Gaussian ansatz, Eq.~(\ref{eq:ansatz}).  The differential equations for $(Q,P,u,v)$ in the semiclassical limit are now given by
\begin{equation} \label{eq:qpsc}
\dot{Q} = \frac{P}{m}	\,\,\,\,\,\,\,\,\,\,\,\,\, \dot{P} = - U_o'(Q) 
\end{equation}
\begin{equation} \label{eq:uvsc}
\dot{u} = - i \omega'(Q) + U_o'' v	\,\,\,\,\,\,\,\,\,\,\,\,\, \dot{v} = -\frac{1}{m} u.	
\end{equation}

The absence of the $v$-dependent Moyal correction term proportional to $\hbar^2$ in Eq.~(\ref{eq:qpsc}) governing the evolution of $P(t)$ in the semiclassical limit leads to a further separation of the $(Q,P)$ equations of motion, resulting in the center of the Gaussian in Eq.~(\ref{eq:ansatz}) now following a \emph{single} classical trajectory in phase space that corresponds to harmonic motion on the \emph{average} potential $U_o(q)$:  
\begin{equation}Q(t) = Q_o + [Q(0) - Q_o] \cos(\Omega_o t) + \frac{P(0)}{m \Omega_o} \sin(\Omega_o t), 
\end{equation}
where $\Omega^2_o = (\Omega_1^2+\Omega_2^2)/2$ and $Q_o$ are the average squared harmonic frequency and minimum of $U_o(q)$, respectively. $P(t)$ follows the  corresponding average motion equation for momentum.   

\begin{figure}
\centering
  \includegraphics[height=9cm]{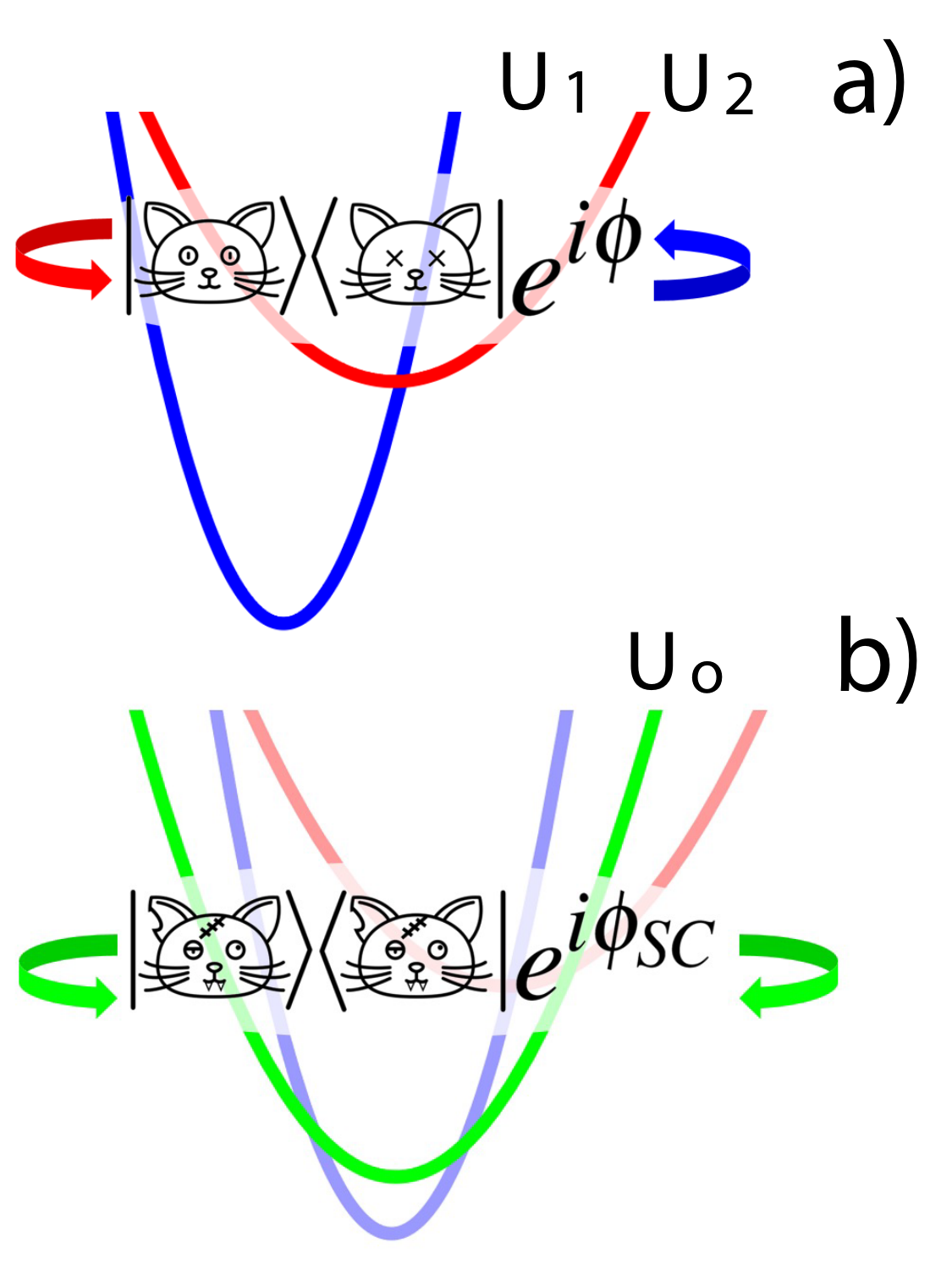}
  \caption{Schr{\"o}dinger's  cat perspective of two state coherence. a) Exact Wigner-Moyal representation, with cats simultaneously alive and dead.  b) Semiclassical approximation with a single average \emph{zombie} state, somewhere between alive and dead, evolving on a single average potential. See text for discussion.}
  \label{fig:fig1}
\end{figure}

We next consider the semiclassical-limit solutions of the $(u,v)$ equations.  These are not independent of $(Q,P)$ but are driven by their dynamics in the form of an external time-dependent force.  The two first-order ordinary differential equations of Eq.~(\ref{eq:uvsc}) can be rewritten as a single second order equation:
\begin{equation} \label{eq:ode2}
\ddot{u} + \Omega_o^2 u = - i \omega'' \dot{Q}(t)	
\end{equation}
This is the equation of motion for a harmonic oscillator $u(t)$ with frequency $\Omega_o$ \emph{resonantly driven} by a sinusoidal perturbation $\dot{Q}(t)$, also with frequency $\Omega_o$.  This leads to secular growth of the parameters $(u,v)$ with time, with terms like $t \sin(\Omega_o t)$ and $t \cos(\Omega_o t)$ showing up in their time dependence. The semiclassical coherence evolution is depicted schematically in Fig.~\ref{fig:fig1} b).   

From a Schr{\"o}dinger's cat perspective \cite{schr35,schl01,wine2013}, the evolution of the exact quantum coherence $\rho_{12}$ in the Wigner-Moyal representation is guided by the simultaneous contributions of a ``dead'' cat trajectory on state 1 and a ``live'' cat trajectory on state 2.  The phase space center of the coherence Gaussian evolves along a phase space path given by the arithmetic mean of the \emph{classical} live and dead motions, a \emph{nonclassical} phase space path that is tempting to call a \emph{cat-jectory}.  

The purely imaginary parameters $u$ and $v$ depend on the difference between classical state 1 and 2 trajectories.  As indicated by Eq.~(\ref{eq:ansatz}), these terms determine the phase space oscillatory structure of the function $\rho_{12}(q,p,t)$. Importantly, the magnitudes of $u(t)$ and $v(t)$ as described by the exact equations, Eqs.~(\ref{eq:uvdex})--(\ref{eq:pj}), are bounded as functions of time, leading to a stable oscillatory behavior of $\rho_{12}(q,p)$ as a function of $q$ and $p$ in phase space.

In the semiclassical limit, two serious pathologies modify the exact quantum behavior.  First, the nonclassical superposition of simultaneous and distinct classical motions corresponding to the ``live'' state 2 and ``dead'' state 1 that make up the guiding cat-jectory and characterize the exact Wigner-Moyal dynamics collapses to a single effective path, neither alive nor dead but somewhere in between---a ``zombie'' state guided by a single classical trajectory on the average potential potential $U_o(q)$. For the exact Wigner-Moyal dynamics, both the mechanical frequencies $\Omega_1$ and $\Omega_2$ appear in the cat-jectory $(Q(t),P(t))$ as well as the other parameters and any observables computed from $\rho_{12}$.  In the semiclassical limit, however, only the single---and nonphysical---average frequency $\Omega_o$ appears, leading to significant errors in the phase space evolution and resulting observables in semiclassical methods.  

A second pathology results from the the qualitative difference in the dynamics of the parameters $(u,v)$ in the exact and semiclassical cases.  For exact dynamics, these purely imaginary quantities are proportional to the differences of the state 1 and state 2 trajectories (see Eq.~(\ref{eq:uvdex}), and are thus bounded at long time.  For the semiclassical case, the secular nature of the $(u,v)$ dynamics causes these terms to grow with time.  This spurious behavior causes $\rho_{12}(q,p,t)$ to be an increasingly oscillatory function of $q$ and $p$ in phase space with increasing time. The increasingly high frequency phase space oscillations of $\rho_{12}$ at long times leads to erroneous cancellations and an effective spurious \emph{decoherence} in the semiclassical case.  (It is tempting to call this effect the "ultraviolet cat-astrophy".)

\section{Numerical Results}
\begin{figure}
\centering
  \includegraphics[height=10cm]{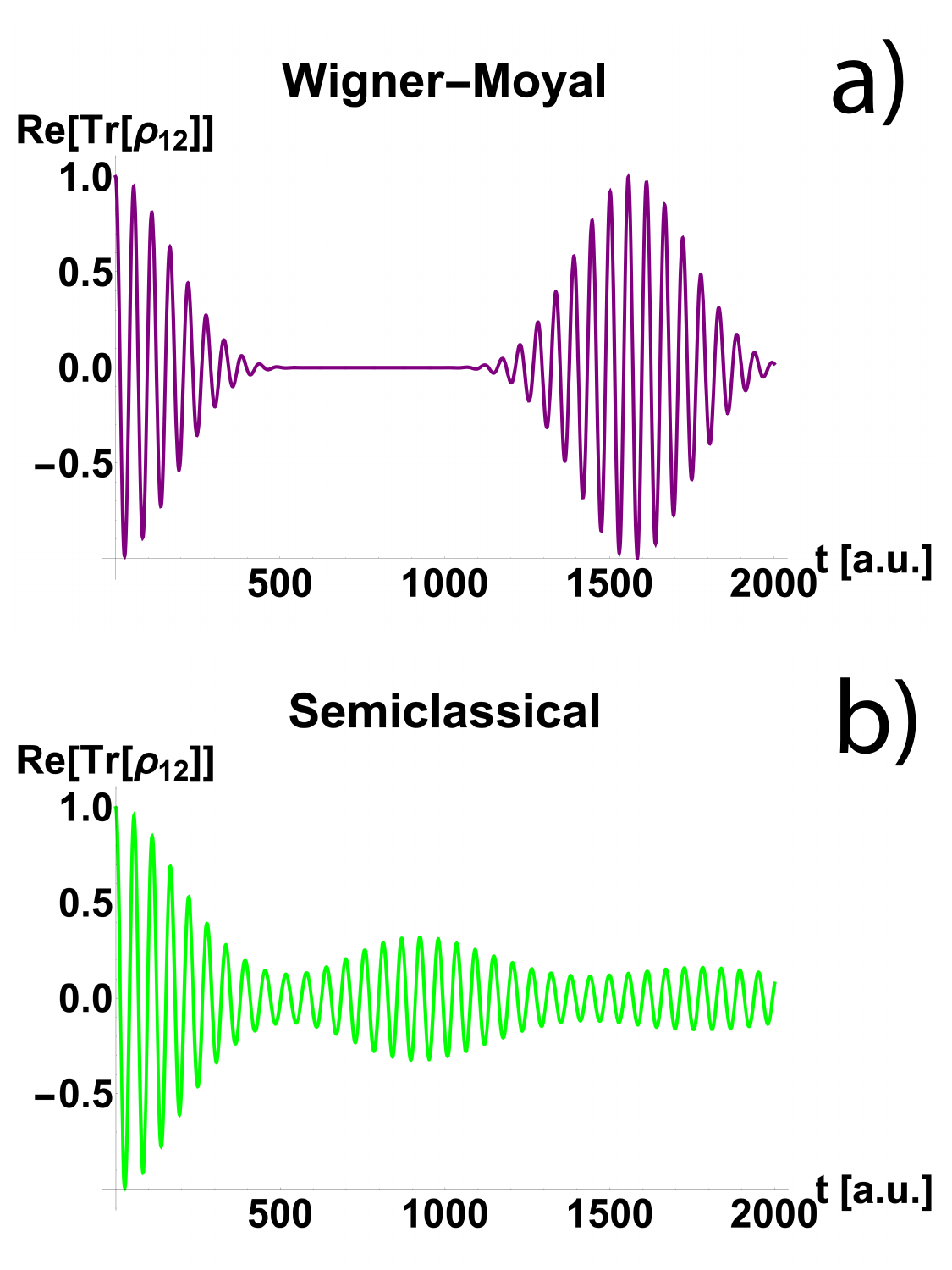}
  \caption{Real part of $c(t)=\mathrm{Tr} \rho_{12}$, for system and initialization described in text. a) Wigner-Moyal (equivalent to exact quantum).  b)  Semiclassical limit of Wigner-Moyal.}
  \label{fig:fig2}
\end{figure}

\begin{figure}
\centering
  \includegraphics[height=10cm]{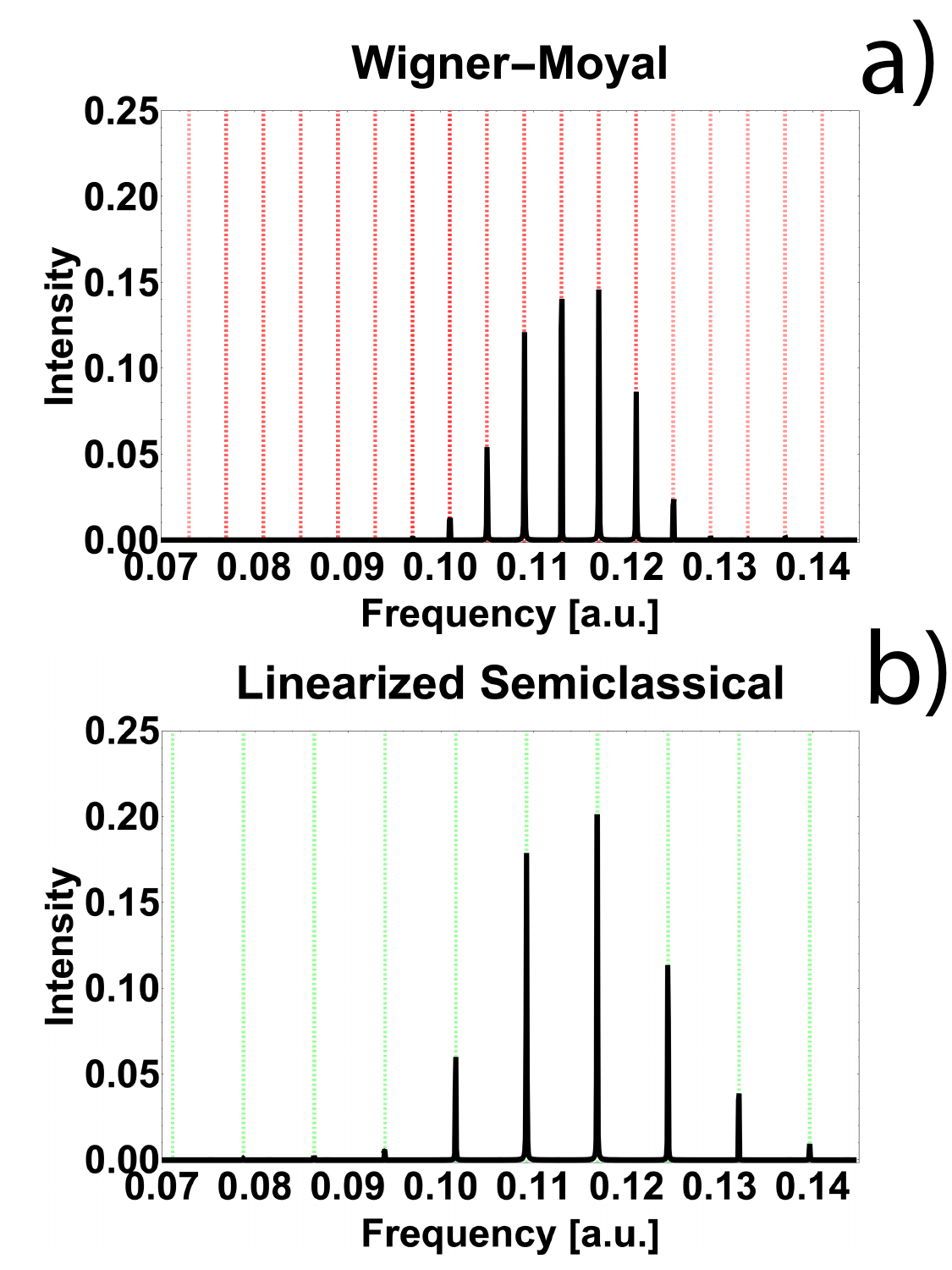}
  \caption{Absorption spectra. a)  Exact Wigner-Moyal.  b) Semiclassical with additional linearized difference potential approximation, as described in text.}
  \label{fig:fig3}
\end{figure}

In Fig.~\ref{fig:fig2} we illustrate the exact Moyal dynamics of the coherence $\rho_{12}$ and its semiclassical approximation by showing the real part of the trace of the coherence $c(t)= \mathrm{Re} \,\mathrm{Tr} \,\rho_{12}(t)$, where the phase space trace $\mathrm{Tr} \,\rho_{12}(t) = \iint \rho_{12}(q,p,t)dq dp$.  With appropriate initial conditions and in the Condon approximation, the Fourier transform of $c(t)$ gives the electronic absorption spectrum of the system \cite{hell18}.  Figure \ref{fig:fig3} shows the corresponding absorption spectra, computed by Fourier transformation of $c(t)$.  

The initial conditions employed correspond to a Gaussian centered at the minimum of the lower (ground) electronic state with zero momentum.  The Gaussian parameters correspond to those of the ground vibrational state Wigner function, and so corresponds to preparing a coherence appropriate for modeling the electronic absorption spectrum of a molecule initially in its ground state. The numerical values of the parameters used are (in atomic units) $m=2000$, $\Omega_1=0.01$, $\Omega_2=0.004$, $Q_1^e=0$, $Q_2^e=1.0$, $E_1^e=0.0$, and $E_2^e=0.1$. 

In Fig.~\ref{fig:fig2} a), we show the exact correlation function in the Wigner-Moyal representation.  The result is a periodic function with recurrences at the period of the upper harmonic surface $T=2\pi/\Omega_2$, which is displaced from its equilibrium by the initial conditions, modulated by the high frequency oscillations corresponding to the electronic energy difference $\omega = (E_1^e-E_2^e)/\hbar$. The Fourier transform of the correlation function gives the absorption spectrum for this system, and is shown in Fig.~\ref{fig:fig3} a).  Here, we see an envelope of spectral lines spaced by the upper state frequency $\Omega_2=0.004$ centered around the electronic transition frequency $\omega = 0.1$, which is the expected absorption spectrum for this system. 

The semiclassical correlation function is shown in Fig.~\ref{fig:fig2} b). Here, the envelope of $c(t)$ does not show a rapid drop to zero with recurrences, but rather an overall decay with the vestiges of the rephasing barely visible.  The features are due to the qualitative differences in the semiclassical parameters compared to the exact Wigner-Moyal case.  The phase space trajectory $(Q(t),P(t))$ in the semiclassical limit does not capture the rapid drop in coherence, complete by $t=500$ in the exact case.  Further, the highly oscillatory structure of $\rho_{12}$ in phase space leads to spurious cancellation in the trace integral defining the trace, and thus an artificial decoherence-like decay.  

The correlation function $c(t)$ in the semiclassical case does not produce an acceptable spectrum due to its spurious decay.  To salvage a semiclassical result for the spectrum, a further approximation can be made by linearizing the difference potential in the computation of the $(u,v)$ dynamics (equivalent to setting $\omega''=0$ in Eq.~(\ref{eq:ode2})).  This eliminates the secular behavior and resulting artificial decay.  The resulting correlation function is now periodic (not shown here).  The spectrum obtained by Fourier transformation of the linearized difference potential semiclassical $c(t)$ is displayed in Fig.~\ref{fig:fig3} b). The spectrum now resembles qualitatively the exact result, with an envelope of sharp lines.  However, the spacing between the lines is now observed to be the incorrect average ``zombie'' frequency $\Omega_o = \sqrt{(0.01^2 + 0.004^2)/2}=0.0076$, rather than the correct upper state frequency $\Omega_2 = 0.004$.  

\section{Summary and Discussion}

In this paper, we have investigated the relation between exact and semiclassical limit dynamics of quantum coherence, represented by the off-diagonal element $\hat{\rho}_{12}$ of the density matrix, for a simple two state model system using the phase space Wigner-Moyal formalism. This approach allows an analytic description of the evolution of $\rho_{12}(q,p,t)$ and clearly highlights pathologies of the commonly employed semiclassical approximation.  The inclusion of the first neglected term in the $\hbar$ expansion of the noncommutative Moyal product restores the full quantum behavior for the harmonic system treated. From the perspective of quantum mechanics provided by Schr\"odinger's famous cat, the corrections to the semiclassical theory resurrects the manifestly quantum simultaneous live \emph{and} dead cat state from an averaged live/dead ``zombie'' behavior that afflicts the semiclassical limit.  

The evolution of a quantum state represented by a Wigner function $\rho_W(q,p,t)$ on a single potential surface has a well-defined classical limit: the evolution of a classical ensemble of trajectories representing a phase space probability density $\rho(q,p,t)$. In the field of chemistry, the Born-Oppenheimer approximation separates the highly quantum mechanical electronic dynamics from the often nearly classical properties of the heavy nuclei, allowing this well-defined  limit to be exploited by using classical methods such as molecular dynamics \cite{alle87} to simulate the dynamics of molecules.

 Simulating molecular systems on the quantum-classical frontier, where quantum electronic transitions accompany nearly classical nuclear dynamics, presents conceptual hurdles and numerical challenges. 
Developing methods based on generalizing classical molecular dynamics to this situation is an active area of research. 
In this paper, we illustrated that nonclassical behavior can appear even in a surprisingly simple system comprised of multi-state dynamics on harmonic potentials with different frequencies. By studying the system analytically, we reveal the nonclassical effects at work, and demonstrate how a naive semiclassical treatment must be generalized to incorporate the nonclassical effects in a classical mechanical framework.  
%The exact results coming from this analytical treatment suggest the basis of further, more approximate numerical methods that incorporate the effects of essential higher-order terms of the Moyal series governing the exact quantum evolution of the quantum coherence---off-diagonal density matrix elements in the Wigner representation.  

These nonclassical effects are experimentally observable, and their inclusion in simulation methodology is important to achieve even qualitative agreement with experiment.  Aside from its fundamental interest, the approach employed here provides insight that may guide the development of such methodologies for incorporating these corrections in the simulation of more realistic complex systems.  This will be discussed in future work.

\section{Acknowledgements}

We thank Greg Ezra and Shaul Mukamel for helpful input and acknowledge stimulating discussions at the UCI Liquid Theory Lunch (LTL).  This work was supported by the National Science Foundation under grant CHE-1764209.

\bibliography{zombiecat}
\end{document}